\documentclass[aps,12pt,a4paper]{revtex4-2}

\usepackage{epsfig}
\usepackage{graphicx}
\usepackage{amsmath,amssymb,color}
\usepackage[english]{babel}
\usepackage{natbib}
\usepackage{appendix}
\usepackage{caption}
\usepackage{subcaption}

\newcommand{\be}{\begin{equation}}
\newcommand{\ee}{\end{equation}}
\newcommand{\eq}[1]{(\ref{#1})}
\newcommand{\fig}[1]{Fig.~\ref{#1}}

\newcommand{\la}{\left<}
\newcommand{\ra}{\right>}
\newcommand{\eps}{\varepsilon}

\begin{document}

\title{Math behind everyday life: ``black days'', their manifestation as traffic jams, and beyond}

\author{Daniil Fedotov$^1$ and Sergei Nechaev$^2$}

\affiliation{$^1$UFR Sciences, Universit\'e Paris Saclay, 91405 Orsay Cedex, France \\ 
$^2$LPTMS, CNRS -- Universit\'e Paris Saclay, 91405 Orsay Cedex, France}


\begin{abstract}

In our daily lives, we encounter numerous independent events, each occurring with varying probabilities over time. This letter delves into the scientific background behind the inhomogeneous distribution of these events over time, often resulting in what we refer to as ``black days'', where multiple events seem to converge at once. In the first part of the work we performed an analysis involving $D$ independent periodic and random sequences of events. Using the Uniform Manifold Approximation and Projection (UMAP) technique, we observed a clustering of event sequences on a two-dimensional manifold ${\cal M}$ at a certain large $D$. We interpret this clustering as a signature of ``black days'', which bears a clear resemblance to traffic jams in vehicle flow. In the second part of the work we examined in detail clustering patterns of independently distributed $N$ points within the corners of a $D$-dimensional cube when $1\ll N<D$. Our findings revealed that a transition to a single-component cluster occurs at a critical dimensionality, $D_{cr}$, via a nearly third-order phase transition. Analyzing the spectral density, $\rho(\lambda)$, of the corresponding adjacency graph in the vicinity of the clustering transition we recover the singular ''Lifshitz tail'' behavior at the spectral boundary of $\rho(\lambda)$.

\end{abstract}

\maketitle

\section{Introduction}
\label{sect:1}

People are accustomed to the fact that events in everyday life are distributed highly nonuniformly: among days that are more or less equiprobably filled with routine activities, there are ``black days'' when it seems that ``everything is happening at once''. Without delving deeply into reasons for such nonuniformity, we tend to interpret it as ``fate''. This fatalistic attitude to the distribution of events in life is essentially a defensive mechanism that allows us to reconcile and accept the inevitability of ``black days''.

Is the presence of black days a subjective feeling of the individual, or are there objective mathematical laws behind it? Could we control the distribution of black days and to some extent predict their occurrence? Could we propose a mathematical model within which we would be able to explore the laws of the emergence of black days? Asking such questions is crucial: by seeking a logical explanation for what we were previously willing to accept, we are turning from passive observers of our own lives into active individuals trying to manage our time and control the events that happen.

In this paper, we describe a rather crude model that demonstrates certain characteristics of distributions typical for the emergence of black days. Our everyday life is full of independent events which happen with some probability over some time period. To have an example, these events could be as follows: (1) school meeting, (2) visit to the doctor, (3) conference deadline, (4) shoes repair, (5) furniture purchase, (6) work report, (7) ``burning'' theater tickets, (8) relative's illness, (9) visit to the bank, (10) bathroom faucet malfunction, (11) waiting for the postman, (12) guest visit, (13) lecture reading, (14) working with a student, (15) business trip, etc... If these (or other) events happen sequentially, we are able to manage them easily, however if too many events coincide in time, we are getting stacked and say that the ``black day'' has arrived.

Below we formulate a model of sudden accumulation of different events which possesses the critical emergence of black days. Each event can be considered to occur in time independently of the others. Therefore, when dealing with, for example, $D$ events, we are essentially distributing points in a $D$-dimensional space, where each axis represents a specific event. By random placing points in this $D$-dimensional space with an independent uniform distributions along the coordinate axes, we explore the full $D$-dimensional space. It is reasonable to anticipate that at small or moderate values of $D$, the points cover the $D$-dimensional space relatively uniformly. However, as $D$ becomes significantly larger, the points tend to form clusters of ``dense spots'' that are distributed non-uniformly in space. These clusters can be associated with black days. The effect of clustering has been discussed in \cite{Beyer1999} and is known as the ``dimensionality curse''. At length of our work we are focused on the analysis of statistical properties of these clusters as a function of $D$ and other microscopic parameters as well.

The concept of ``black days'' yet seems too abstract to be directly compared with any model. However, the spontaneous accumulation of vehicles which happens in traffic flows provides a clear, quantifiable example of a ``black day''. At low vehicle densities the exit rate, $q$, is proportional to the vehicle flow density. This relationship holds until the exit rate reaches its maximum, $q_{max}$. Beyond this point, the exit rate begins to decline. Such a behavior aligns with the accumulation of independent events: the vehicle density can be seen as the (normalized) number of independent events, and the exit rate, $q$, -- as the number of events manageable within a short time frame. From this perspective, the maximum exit rate corresponds to the highest density of manageable events, while the subsequent decline signifies the onset of a ``black day''. This ``traffic jam'' interpretation of a ``black day'' we will keep in mind when discussing the accumulation of sequences of events.

The structure of our work can be outlined as follows. In Section \ref{sect:2}, we present a model describing the simultaneous distribution of many independent events over a considerable time period. By analyzing the corresponding time sequences using the UMAP (Uniform Manifold Approximation and Projection) technique, we illustrate that the clustering of data is contingent upon the number of sequences. In the same Section we also provide the comparison of our results with various scenarios of traffic flows. Section \ref{sect:3} deals with the detailed investigation of properties of ensembles of independently distributed points in a $D$-dimensional space where we examine basic statistical characteristics of forming clusters. In Section \ref{sect:4} we consolidate our findings and engage in a discussion regarding potential future developments. In Appendix \ref{app1}, we apply the ``Laplacian Eigenmap'' technique to the analysis of an ensemble of time series, demonstrate the clustering of data in a high-dimensional space and designate a class of problems where this technique may offer advantages over UMAP, while in Appendix \ref{app2} we discuss the distinction between clustering in random regular graphs and in sparse matrices near the percolation threshold. 

\section{Clustering of multiple independent periodic sequences}
\label{sect:2}

\subsection{The model}
\label{sect:2.1}

Encountering ``black days'' we can model using a following simplistic framework. Consider a significant duration of time, denoted as $T$, such as, say, one year. Within this time span, various independent activities take place. We pay attention to three types of activities: (i) periodic activities, each characterized by its unique period and initial phase, and (ii) quasiperiodic activities, and (iii) random activities.  

1. The periodic activities are constructed as follows. Suppose that we have $D$ independent sequences of events. Each of them we describe by a spiking time-series $x_{j}(t)$ ($j=1,...,D$): 
\be
x_j(t) = \frac{\delta^2}{\left(\sin(a_j t+\phi_j)-1\right)^2+\delta^2}
\label{eq:01}
\ee
Here, time $t$ is considered continuous, it is changing within the interval $0\leq t \leq T$. The parameters $a_j$ and $\phi_j$ in \eq{eq:01} represent respectively the frequency and initial phase of activity $j$. Throughout our analysis, we treat $a_j$ and $\phi_j$ as uniformly distributed random variables. The parameter $\delta$ ($0<\delta\ll 1$), controls the sharpness of spikes, essentially setting the resolution threshold of our model. It delineates the minimal time scale within which two distinct events (spikes) from different sequences can be significantly overlapped.

\begin{figure}
\includegraphics[width=0.8\linewidth]{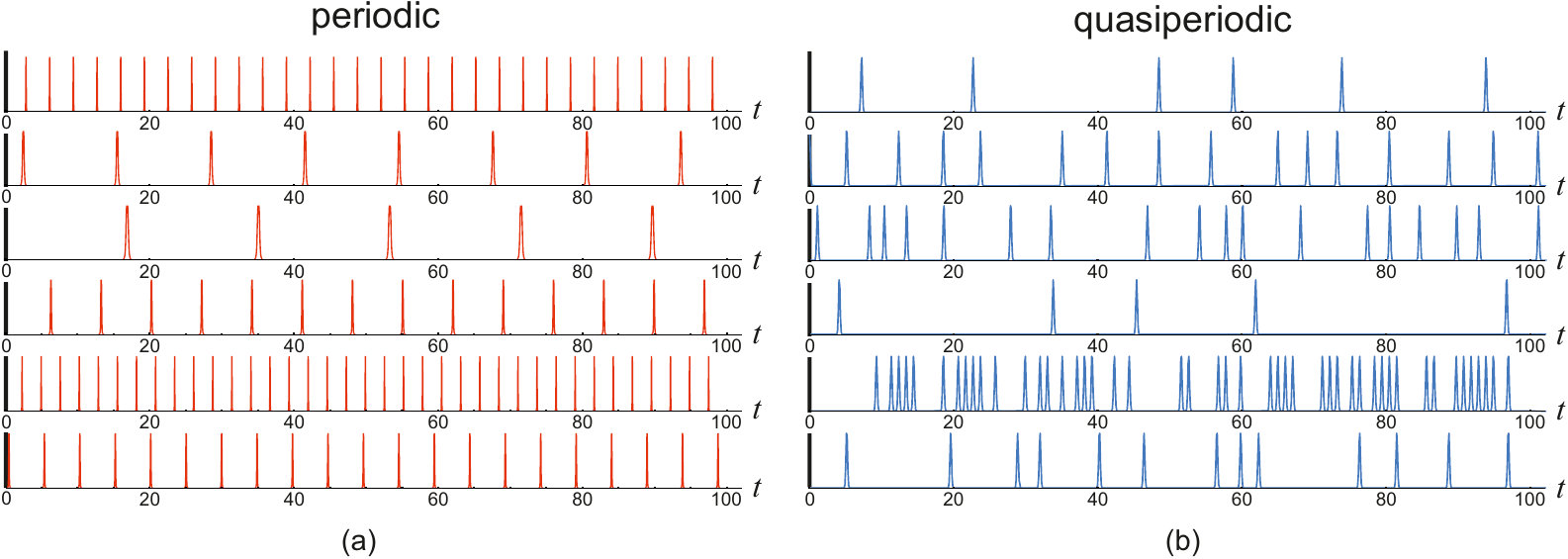}
\caption{Sample plots of $D=6$ independent spiked activities up to ``time'' $T=100$: (a) periodic activities with random phases and periods; (b) quasiperiodic activities obtained from periodic ones by ``smeariong'' the positions of spikes with the Gaussian distribution characterized by the dispersion $\sigma=2$.}
\label{fig:01}
\end{figure}

We generate $D$ time series $\left\{x_1(t), x_2(t), \ldots, x_D(t)\right\}$ by assigning each series $j$ ($1\le j \le D)$ its own values $a_j$ and $\phi_j$, which are drawn from uniform distributions: $P(a_j) = 1/\Delta_{a}$ where $\Delta_{a} = T = 100$, and $P(\phi_j)=1/a_j$. That is, for the sequence $j$ we randomly generate a period $a_j$ uniformly distributed within the interval $[0,T]=[0,100]$, and then for a generated period $a_j$ we randomly generate a phase $\phi_j$ which is uniformly distributed within the interval $[0,a_j]$. The sample plots of $D=6$ such independent periodic activities up to $T=100$ are shown in \fig{fig:01}a. 

2. To generate a set of quasiperiodic activities we take a periodic activity with a period, $a_j$, and shift randomly each spike away from its prescribed position, $x(t)$, according to the Gaussian distribution with the dispersion $\sigma$, i.e. we ``smear'' the periodic distribution of peaks. The sample plots of $D=6$ such sequences of randomised activities up to $T=100$ are shown in \fig{fig:01}b for relatively small value of the dispersion (in our case $\sigma=2$) selected in such a way that the typical shift of the spike from its nonrandom position is less than the typical period of the sequence $\Delta_a$.

3. The set of random activities (not shown in \fig{fig:01}) is generated in the same way as the set of quasiperiodic ones, however with a larger dispersion $\sigma$, such that the information about the prescribed position of a spike in an initial periodic sequence is fully washed out. 

To visualise the ``condensation'' of activities depicted by independently generated sequences, we employ a dimensional reduction technique. We observe the transition at the path level, using the nonlinear dimension reduction algorithm UMAP (Uniform Manifold Approximation and Projection) \cite{mcinnes2018umap}. This algorithm projects the high-dimensional dataset onto a two-dimensional manifold ${\cal M}$, where each point on ${\cal M}$ represents a specific time-dependent activity. The emergence of ``black days'' manifests itself as the condensation of points on ${\cal M}$. Briefly, UMAP constructs a graph ${\cal G}$ embedded in the Euclidean $D$-dimensional space, which represents proximity of data by connecting $\kappa$ nearest neighboring points of ${\cal G}$ ($\kappa$ is a parameter of the model), and then optimizes an ``effective'' graph embedded in a 2D manifold ${\cal M}$ to be as structurally similar as possible to ${\cal G}$.

\begin{figure}
\includegraphics[width=0.8\textwidth]{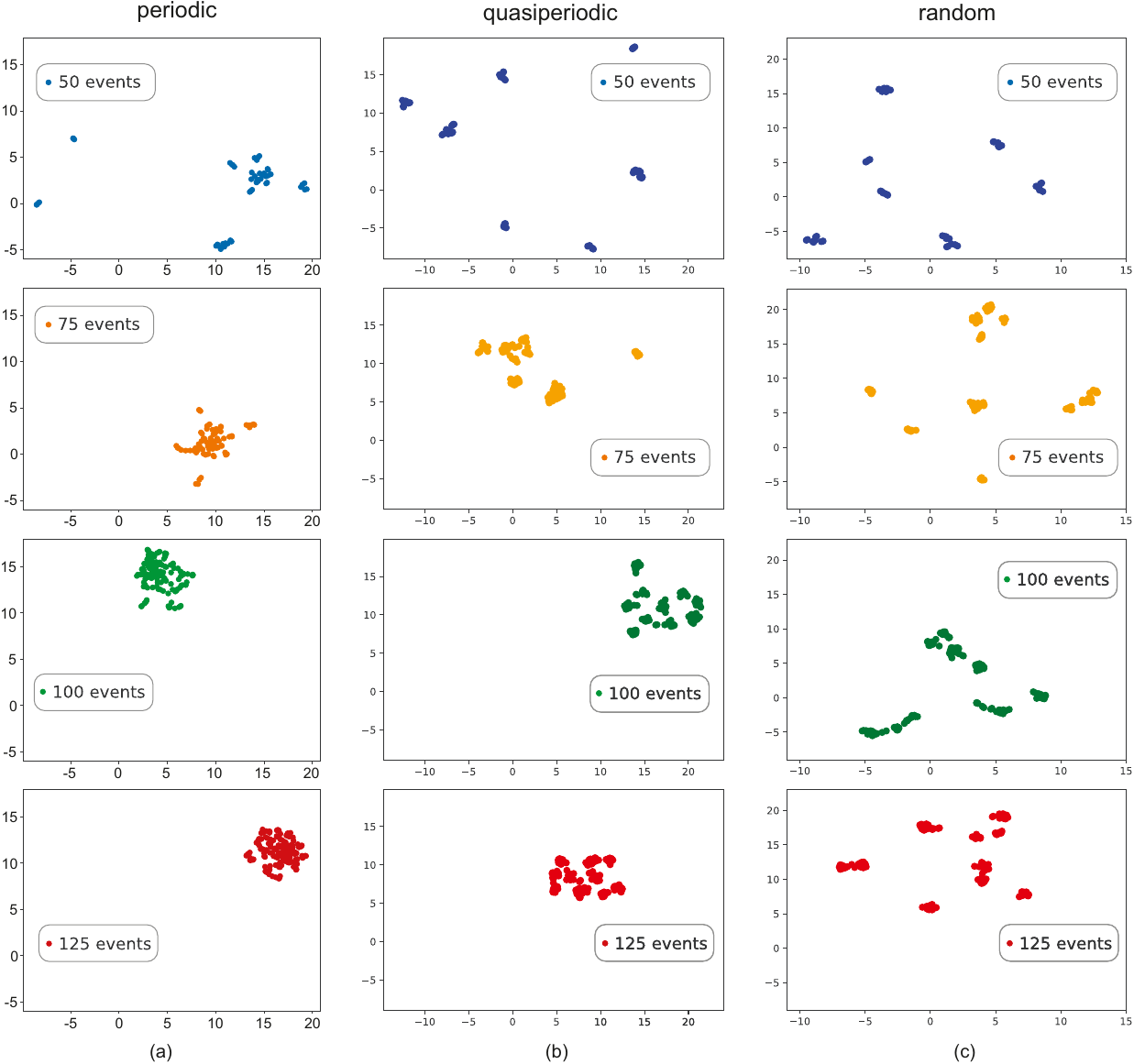}
\caption{Two-dimensional snapshots of activities obtained using UMAP algorithm. The condensation of independent sequences of spikes for $D=\{50, 75, 100, 125\}$ sequences is clearly seen with increasing $D$ for: periodic (a), quasiperiodic (b), and random (c) sequences of activities.}
\label{fig:02}
\end{figure}

For the dataset undergoing dimensional reduction, we construct the ensemble comprising the set of $D$ vectors $\left\{{\bf x}_1, {\bf x}_2,...,{\bf x}_D \right\}$, with each entry ${\bf x}_j$ ($j=1,...,D$) being a vector ${\bf x}_j=(x_j(1),..., x_j(t), ..., x_j(T))$, where $T$ represents the total sequence length. For the ensemble of periodic sequences we vary number of sequences, $D$, within the range $D \in [1, 125]$, while keeping the parameters $\Delta_{a}$ and $\Delta_{\phi}$ fixed and examine snapshots of ensemble of our periodic activities projected onto the two-dimensional manifold ${\cal M}$ for a predefined value $\kappa$. For the set of quasiperiodic and activities we proceed in the same way while keeping fixed the parameter $\sigma$ of the Gaussian distribution. Specifically, we take $\sigma = 2$ and $\sigma = 40$ for quasiperiodic and random sequences respectively. The outcome of our analysis is illustrated in \fig{fig:02}a-c. The similar analysis has been also performed using an alternative approach known as Eigenmap technique (``Laplacian Eigenmaps''). The similarities and differences of UMAP and Lalacian Eigenmaps is discussed in the Appendix.

Above a certain $\kappa$-dependent threshold, $D_{cr}$, the data exhibits a tendency to coalesce into a dense cluster, indicating the emergence of a ``black day''. In the following Section, we provide the analysis of the clustering in the transitional region in the vicinity of $D_{cr}$.

To characterize quantitatively the sparseness/condensation of points in the two-dimensional UMAP plot we introduce the natural parameter, $s$, which designates the area, $S$, of the minimal rectangle encompassing all points on the two-dimensional UMAP manifold ${\cal M}$, divided by the number of sequences, $D$, i.e., $s=S/D$. Dimensional normalization ensures that the growth of the data set does not distort the analysis. This allows data to be compared and analyzed consistently on a common density scale in the projected space.

\begin{figure}[ht]
\includegraphics[width=0.8\linewidth]{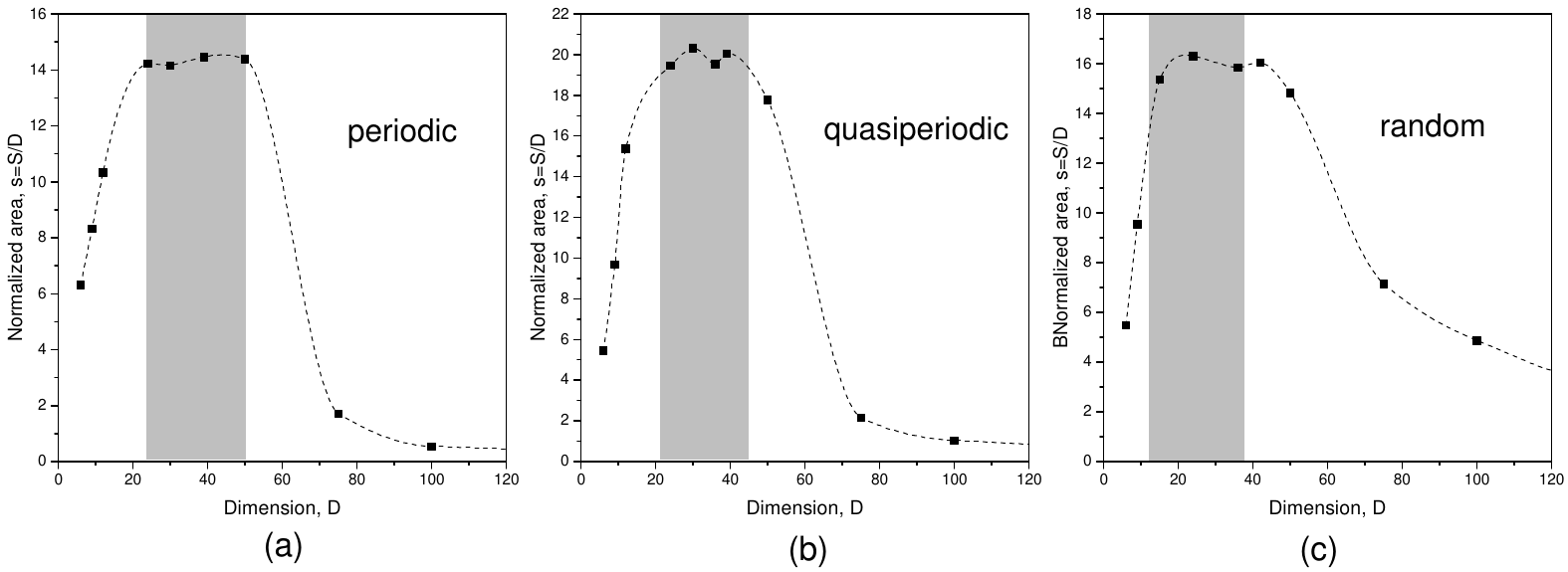}
\caption{Nonmonotonic dependence of the normalized area, $s=S/D$, of a minimal rectangle surrounding UMAP points (see \fig{fig:02}) on the number of sequences, $D$, for $D/\kappa=25$: (a) periodic, (b) quasiperiodic, (c) random.}
\label{fig:03}
\end{figure}

The plots $s(D)$ are drawn in \fig{fig:03}a-c in the wide range of dimensions $D$ ($D\in [1,120]$) for periodic (\fig{fig:03}a), quasiperiodic (\fig{fig:03}b) and random (\fig{fig:03}c) sequences of activities at fixed quotient $D/\kappa$. The condition $D/\kappa=\mathrm{const}$ has natural meaning since the more activities one considers during a total time, $T$, the higher the expected number of these activities occurs within a specific overlapping period (referred to as the ``day''). To account for this phenomenon, we maintain the ratio $D/\kappa$ constant.

Interestingly, the behavior of the normalized area $s(D)$ is nonmonotonic and for a fixed set of parameters grows, then exhibits a plateau, and finally drops down at $D_{cr}$. The origin of the nonmotonicity could be understood from the following arguments. At small $D$ the points on a two-dimensional UMAP manifold ${\cal M}$ form a dense cluster because for a given value of the graph connectivity, $\kappa$, many points in the graph ${\cal G}$ are connected and hence ${\cal G}$ is highly clustered. At intermediate $D$ the graph ${\cal G}$ becomes relatively sparse since the existing $\kappa$ gets smaller than the typical vertex degree of ${\cal G}$ if it would be considered as a complete graph. At large $D$ the $2D$ UMAP projection represents again a dense spot due to clustering of vertices of ${\cal G}$ in a high-dimensional space. Below we provide possible interpretation of this nonmonotonicity in terms of the traffic flow.

\subsection{``Black days'' and traffic jams}
\label{sect:2.2}

The spontaneous accumulation of events in multiple independent time series, whether periodic or random, can be directly interpreted as vehicle congestion in traffic flows. In this context, each spike in a given time series represents the moment when a vehicle passes through a specific registration point along its route. Consequently, the spiked time series of events can be mapped onto a series of time intervals, indicating the gaps between a vehicle's visits to successive registration points. The number of distinct time series corresponds to the number of vehicles and is therefore proportional to the traffic flow density, $\rho$, measured in cars per unit time interval.

Traffic flow analysis using UMAP differs in several ways from standard techniques, particularly when traffic is measured at a single point on a road map. As a nonlinear method, UMAP provides insights into the global structure of the manifold ${\cal M}$, which is constructed through a collective analysis of all time series from their starting points to their endpoints. From this perspective, UMAP seems to be particularly useful when we are interested in understanding the overall traffic flow and the global congestion within the area of the road map under analysis.

\begin{figure}[ht]
\includegraphics[width=0.8\linewidth]{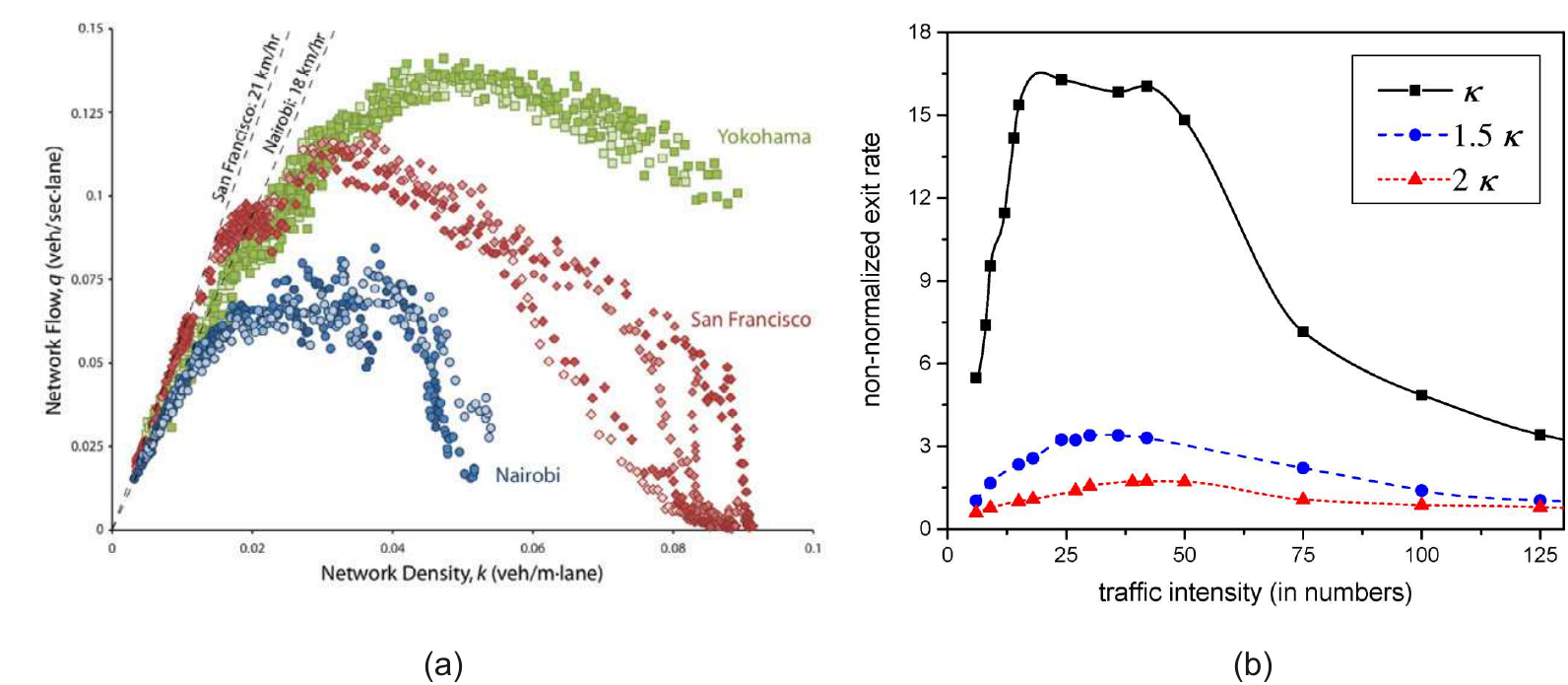}
\caption{(a) Flow of vehicles, $q$, as a function of a vehicle density, $k$ (reproduction of Fig.1d from \cite{traf1}); (b) Dependence of the normalized area, $s=S/D$, of a minimal rectangle surrounding UMAP points (see \fig{fig:02}) on the number of sequences, $D$, for three different values: $D/\kappa=25$ (black-square curve), $D/(1.5 \kappa)=50/3\approx 16.7$ (blue-circle curve), $D/(2 \kappa)=12.5$ (red-triangle curve).}
\label{fig:04}
\end{figure}

Samples of traffic flows in several cities cities are depicted in \fig{fig:04}a, reproduced from the work \cite{traf1} and \cite{traf2}. Figure \ref{fig:04}a presents the results of real measurements, showing the exit rate of cars, $q$, as a function of car density, $k$. One sees that the topology of the environment essentially affects the $q(k)$ relationship.

Let us qualitatively compare \fig{fig:04}a with \fig{fig:04}b, which presents the same results as \fig{fig:03}c for UMAP analysis of abstract multiple random spiked sequences. The three different plots in \fig{fig:04}b correspond to three distinct thresholds $D/\kappa$. Remind that $D/\kappa$ is a parameter that controls the degree of the vertices of the graph ${\cal G}$ in the multidimensional space (see the explanations at length of Section \ref{sect:2.1}). This parameter has a clear interpretation in the context of traffic flow, characterizing the effective interaction between vehicles and thus could reflect the typical road accessibility. When $D/\kappa$ is large, the vertex degree of the graph ${\cal G}$ is relatively small, the interactions are weak (indicating broad roads) and traffic flows smoothly, reaching high values as shown by the black-square curve in \fig{fig:04}b. Conversely, when $D/\kappa$ is small, the vertex degree of the graph ${\cal G}$ is large, the interactions between vehicles are strong (indicating narrow roads), and the vehicle flow increases more slowly, reaching lower absolute values of the exit rate -- see the blue-circle and red-triangle curves in \fig{fig:04}b. This prediction based on the UMAP analysis of an abstract model rhymes with real traffic flow measurements reproduced in \fig{fig:04}a. 

Typically, traffic flow models assume vehicle ordering and incorporate specific local interactions (\cite{chowdhury} for review). The typical examples are the TASEP (Totally Asymmetric Simple Exclusion Process)-like model \cite{derrida} described by the KPZ (Kardar-Parisi-Zhang) equation in continuous limit, or generalized TASEP model proposed by Nagel and Schreckenberg in \cite{nagel}. Our approach, based on the UMAP technique, involves a cumulative analysis of time series, where the primary information about vehicle interactions is encoded in the intervals between spikes. When comparing our \fig{fig:04}b with Fig. 1 in the paper by J. de Gier et al. \cite{gier}, a striking similarity between the corresponding density-flow diagrams becomes apparent. This raises an interesting open question: does the UMAP approach reproduce the KPZ scaling \cite{gier} found in TASEP-like models?

\section{Clustering of points in a high-dimensional cube}  
\label{sect:3}

The condensation of sequences above a dimension $D_{cr}$ suggests delving deeper into the structure of emerging clusters. Let us forget about the sequences and construct a $D$-dimensional vector ${\bf y} = (y_1, y_2, ..., y_D)$, where each component $y_k$ ($k=1,...,D$) acquires random values 0 or 1 with equal probabilities. Considering the connection with the model discussed in the previous Section, it is informative to interpret "1" as indicating the occurrence of a specific event on day $t$, while "0" signifies its absence on the same day. We can think of this collection of $D$ independently generated ones and zeros as mimicking a series of events happening on a particular day, $t$, effectively representing a ``section'' of a multitude of sequences $x_k(t)$. The vector ${\bf y}$ corresponds to a corner in a $D$-dimensional cube with an edge length of 1. 

Generate a set of $N$ independent random vectors $\{{\bf y}^{(1)}, {\bf y}^{(2)},...,{\bf y}^{(N)}\}$ and construct a proximity $\eps$-graph, ${\cal G}$, as follows: the graph ${\cal G}$ has an edge if the normalized by $D$ Euclidean distance $d_{ij}$ between two points ${\bf y}^{(j)}$ and ${\bf y}^{(k)}$ does not exceed ${\eps}$, meaning that 
\be
d_{ij} = \frac{1}{D}\sqrt{\sum_{k=1}^D{\left(y_k^{(i)}-y_k^{(j)}\right)^2}}\le \eps\quad \forall (i,j) \in [1, N]
\label{eq:02}
\ee
Graphs constructed in this manner are known as ``Random Geometric Graphs'' (RGGs), and their geometric and statistical properties, particularly the percolation transition, have been extensively discussed in the literature. Without attempting to provide an exhaustive list of references, we highlight the works \cite{erdos-spenser, Ajtai, Bollobs1983, Borgs1, Borgs2, Borgs3, Bhamidi, Metz, Serrano} that focus on various statistical and geometric aspects of RGGs, their finite-size effects near the percolation transition, and the links between RGGs and Erd\H{o}s-R\'{e}nyi graphs.

Here we are interested in how the average number of disjoint components of the graph ${\cal G}$ depends on the dimensionality, $D$, given the fixed proximity threshold $\eps$. Specifically, we seek the critical $D_{cr}$ at which the entire graph comprises a single component. Recall that the number of disjoint components, $r$, of a graph ${\cal G}$ is determined by the degeneracy of the zero eigenvalue of the graph Laplacian, $L=\hat{D}-A$, where $\hat{D}$ is the diagonal matrix of all vertex degrees of the graph ${\cal G}$ and $A$ is the adjacency matrix of ${\cal G}$. We examine the structure of the graph ${\cal G}$ at $D_{cr}$ and above $D_{cr}$ by analyzing the spectral density of the adjacency matrix $A$. We discuss the phase transition at $D_{cr}$, characterized by a critical exponent $\nu$, defined as $\omega_r \sim N^{-1\nu}$, where $\omega_r$ represents the transition width of an $N$-vertex graph at $r=1$ when the whole network shrinks to a single-component graph i.e. when the zero eigenvalue of the graph Laplacian, $L$, becomes singly-degenerated.

Let us emphasize that we are considering the situation in which the number $N$ of vertices in the graph ${\cal G}$ is always less then the dimensionality $D$ of the space in which ${\cal G}$ is embedded, i.e. $N<D$. The opposite regime $N\gg D$ is not informative since each vertex of the $D$-dimensional cube contains many graph edges. This leads to an essential degeneration of all structural properties of the graph. Specifically, the numerical simulations discussed below run for $N=25...100$ and $D=150...500$.

\subsection{Condensation of components of ${\cal G}$ in a $D$-dimensional cube}
\label{sect:3a}

To determine numerically the order of the condensation transition of ${\cal G}$ into a unique-component graph, we proceed as follows. Let ${\cal G}$ be a graph with $N$ vertices and $r$ disjoint components embedded in a $D$-dimensional cube. Select $r$ as the order parameter, and define $D_{cr}$ as the dimensionality of the space at which $r=1$ when the graph size $N$ tends to infinity. The deviation of the space dimensionality, $D$, from the critical value $D_{cr}$ is the control parameter, $h = D-D_{cr}$. 

The phase transition is well defined in the thermodynamic limit $N\to\infty$ only. It is well known \cite{binder} that for a finite system the order of the phase transition, $\nu$, can be extracted from the finite-size dependence of the transition width, $\omega_1$ on $N$ in the vicinity of the transition point, $D_{cr}(N)$, i.e. at $h=0$. Following the ideas of the work \cite{binder} we look at the shift of the critical dimensionality $D_{cr}(\infty)$ as the function of the graph size:
\be
D_{cr}(\infty) - D_{cr}(N) \sim N^{-1/\nu}
\label{eq:critical}
\ee
where $\nu$ is some scaling exponent. The order of the phase transition is then determined as $\theta = \nu$. 

\begin{figure}[ht]
\includegraphics[width=0.8\textwidth]{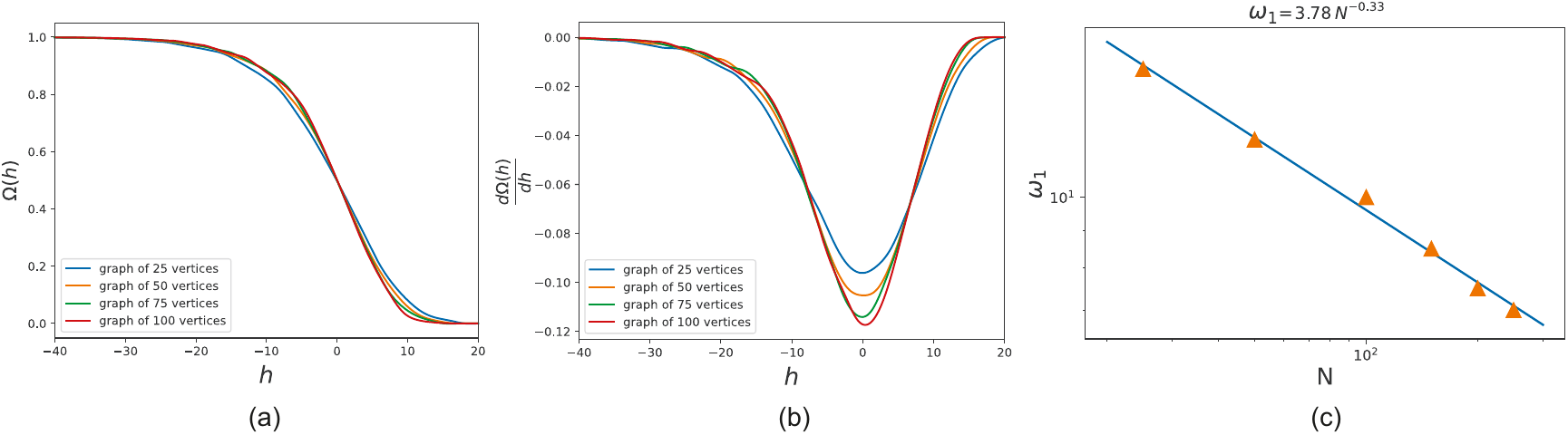}
\caption{(a) Dependence of the normalized number of disjoint graph components, $\Omega_r(h)$, on $h = D-D_{cr}$ for graphs of sizes $N=25,50,75,100$; (b) The transition width, defined by $\omega_{r=1}(h)$ at the level $\min(\omega_{r=1}(h))/\sqrt{2}$ for graphs of sizes $N=25, 50, 75, 100$; (c) The dependence (in a log-log scale) of the transition width as a function of $N$.}
\label{fig:05}
\end{figure}

We have drawn in \fig{fig:05}a the set of curves $\Omega_r(h)$ for different values of $N$, where $\Omega_r(h)$ is the number of disjoint components normalized by the number of the graph vertices $N$, and $h = h(N) = D(N)-D_{cr}(N)$. We are interested in the slope of $\Omega_r(h)$ at the point of intersection of curves at different $N$. Specifically, we follow the method described in \cite{valov} and numerically compute the derivative $\omega_{r=1}(h)=\frac{d\Omega_{r=1}(h)}{dh}$. Then we can extract the phase transition scaling from the width of the function $\omega_1(h)$ at the level $\min(\omega_{r=1}(h))/\sqrt{2}$ computed for each $N$ -- see \fig{fig:05}b.

The function $\omega_1$ should be associated with \eq{eq:critical}: namely if $\omega_1$ shrinks with $N$ as $\omega_1\sim N^{-1/\nu}$, the transition order at $N\to\infty$ is $\theta=\nu$ (see \cite{binder}). The power-law approximation $a N^{-1/\nu}$ of the transition width, $\omega_1(N)$, is plotted in \fig{fig:05}c in the log-log coordinates giving us $1/\nu\approx 0.33$, which permits us to suggest that the transition is of a nearly third order, $\theta \approx 3$.

\begin{figure}[ht]
\includegraphics[width=0.8\textwidth]{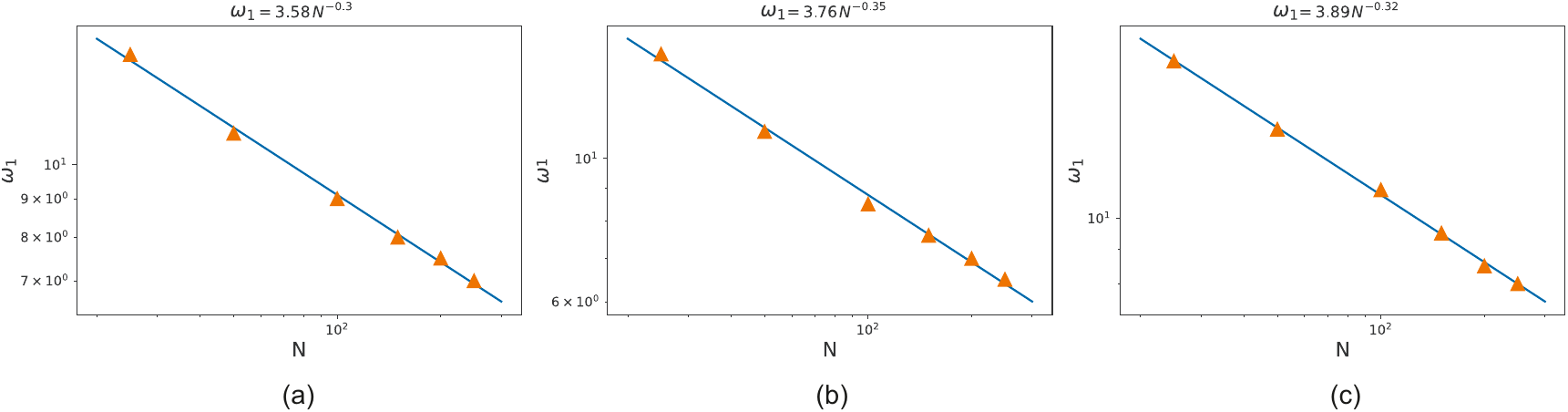}
\caption{The finite-size scaling (in a log-log scale) of the width of the condensation transition as a function of $N$ for the thresholds: (a) $1.5\eps$, (b) $0.8\eps$, (c) $0.5\eps$.}
\label{fig:06}
\end{figure}

As mentioned earlier, the threshold $\eps$ is one of the system parameters. We have repeated all computations for other thresholds taking $\eps'=0.5\;\eps$, $\eps''=0.8\;\eps$, and $\eps'''=1.5\;\eps$, where $\eps=0.0425$ is the threshold used for initial simulations. The resulting values of critical exponents $\nu$ differ by less than $\pm 0.03$, as shown in plots in \fig{fig:06}a-c.

The fact that shrinking of RGG into one-component cluster is the third order phase transition deserves special discussion. In \cite{Borgs1,Borgs2} and later in \cite{Bhamidi} the authors studied random subgraphs of the $D$-cube $\mathbb{Q}_D = \{0,1\}^D$, where nearest-neighbor edges are occupied with the probability $p$. They defined the value $p_c(D)$ for which the expected cluster size of a fixed vertex attains the value $\sim 2^{D/3}$. It has been shown in these works that the largest cluster inside a scaling window of width $\sim 2^{-D/3}$ is of typical size $\sim 2^{2D/3}$. By associating the volume $V = 2^D$ of $\mathbb{Q}_D$ in the works of C. Borgs et al. \cite{Borgs1, Borgs2, Borgs3} with the number of vertices $N$ in our consideration, we can see the consistency of our conclusion regarding the critical exponent $1/\nu = 1/3$ with the results presented in \cite{Borgs1, Borgs2, Borgs3}. 

The fact that single-cluster formation occurs as a third-order phase transition is particularly intriguing, as such transitions typically emerge in the extreme value statistics of correlated many-body systems of the KPZ type \cite{majumdar} or in mean-field-like systems where the transition is driven by geometric factors \cite{baruch,nechaev}. As noted at the end of Section \ref{sect:2.2}, determining whether the results obtained using the UMAP technique are related to those found in many-body systems exhibiting KPZ-type scaling is a challenging open question.

The order of the phase transition depends on the specific choice of the observable, which is conventionally selected as the order parameter. In our case, we have chosen the number of disjoint components, $r$, of the graph in the $D$-dimensional space as the order parameter, and this results in a third-order phase transition. In contrast, in the standard percolation problem for an ensemble of random $N$-vertex Erd\H{o}s-R\'{e}nyi (ER) graphs, one can select the typical size of the connected cluster, $M$, as the order parameter. Below the percolation threshold, $M$ grows as $M \sim N^{2/3}$ (see, for example, \cite{luczak}, as well as \cite{krapivsky} for a review of kinetic approach to statistics of random graphs). The corresponding scaling window, $\Delta$, shrinks as $N$ increases, following $\Delta \sim N^{-1/3}$, which implies $\Delta \sim M^{-1/2}$. This relationship indicates that the phase transition for the quantity $M$ is of second order. 

More intriguing is the condensation of degrees in ER graphs. In \cite{Metz} it was shown that the degree statistics of the classical random graph model undergoes a first order phase transition between a Poisson-like distribution and a condensed phase.

\subsection{Spectral density}
\label{sect:3b}

We have provided more detailed study of the structure of emerging clusters of the RGG ${\cal G}$ when we approach the critical dimension $D_{cr}$ from below by looking at the spectral density of corresponding adjacency matrices. Our reference point for comparison is the spectral density of Erd\H{o}s-R\'{e}nyi graphs in the vicinity of the percolation threshold. ER graph is defined by the symmetric $N\times N$ adjacency matrices $B$, whose matrix elements $b_{ij}$ are:
\be
b_{ij}=b_{ji} = \begin{cases} 
1 & \mbox{with the probability $p$} \\ 0 & \mbox{with the probability $1-p$} 
\end{cases}
\label{eq:03}
\ee
(it is supposed that $b_{ii}=0$ for $i=1,...,N$). Spectrum and topology of ER graphs are controlled by the dependency of $p$ on $N$. Many results are known in cases when $p$ approaches zero slower than $1/N$. Meanwhile, there are many white spots in the case when $p = c/N$, where $c$ is a constant. In \cite{ks2003} Krivelevich and Sudakov proved that for $p\in(0,1)$ the typical largest eigenvalue is 
\be
\lambda_{max} = (1+o(1))\max\{\sqrt{d_{max}},Np\},
\ee
where $d_{max}$ is the maximal vertex degree of ER. In \cite{BenaychGeorges2019, BenaychGeorges2020, Alt2021} other eigenvalues in spectrum are analyzed for different regimes of $p$. Eigenvalues fluctuations are also an object of interest in literature. In recent work \cite{Bhattacharya2021} the lower and the upper large deviations tails of $\lambda_{max}$ were studied for $\frac{1}{N^{o(1)+1}}\ll p \ll \frac{1}{N}\sqrt{\frac{\ln N}{\ln \ln N}}$. In \cite{diaconu2022clts} $\lambda_{max}$ it was stated that the Gaussian fluctuations emerge when $N^{\varepsilon-1} \leq p \leq \frac{1}{2}$, $\varepsilon\in(0,1)$.

\subsubsection{Spectral density of RGG at critical $D_{cr}$} 
\label{sect:3b-1}

The typical structure of the spectral density $\rho_A(\lambda)$ of RGG embedded in a $D$-dimensional cube is depicted in \fig{fig:06}a,b by blue curves at two different critical dimensions $D_{cr}$ for two different graph sizes, $N=200$ and $N=400$. Orange curves in \fig{fig:06}a,b designate spectral densities $\rho_B(\lambda)$ of ensemble of adjacency matrices $B$ of ER graphs in the vicinity of the percolation threshold, $p_{cr}=\tfrac{1}{N}$. One sees from \fig{fig:06} that properly adjusting the control parameters $D$ (for RGG) and $p$ (for ER), it is possible to achieve almost complete coincidence of spectral densities $\rho_A(\lambda)$ and $\rho_B(\lambda)$. According to \cite{rojo1,rojo2}, the maximal eigenvalue of $A$ is bounded by the maximal vertex degree of subgraphs constituting ${\cal G}$. Denoting the maximal vertex degree in the ensemble of subgraphs as $z_{max}$, one has $\lambda_{max} <2\sqrt{z_{max}-1}$ for the largest eigenvalue of $A$. In our simulations carried for $D\lesssim D_{cr}$ we have found the subgraphs with at most $z_{max}=4$, which allows us to estimate $\lambda_{max}$ as $\lambda_{max}\le  2\sqrt{3}\approx 3.46$. 

\begin{figure}[ht]
\includegraphics[width=0.8\textwidth]{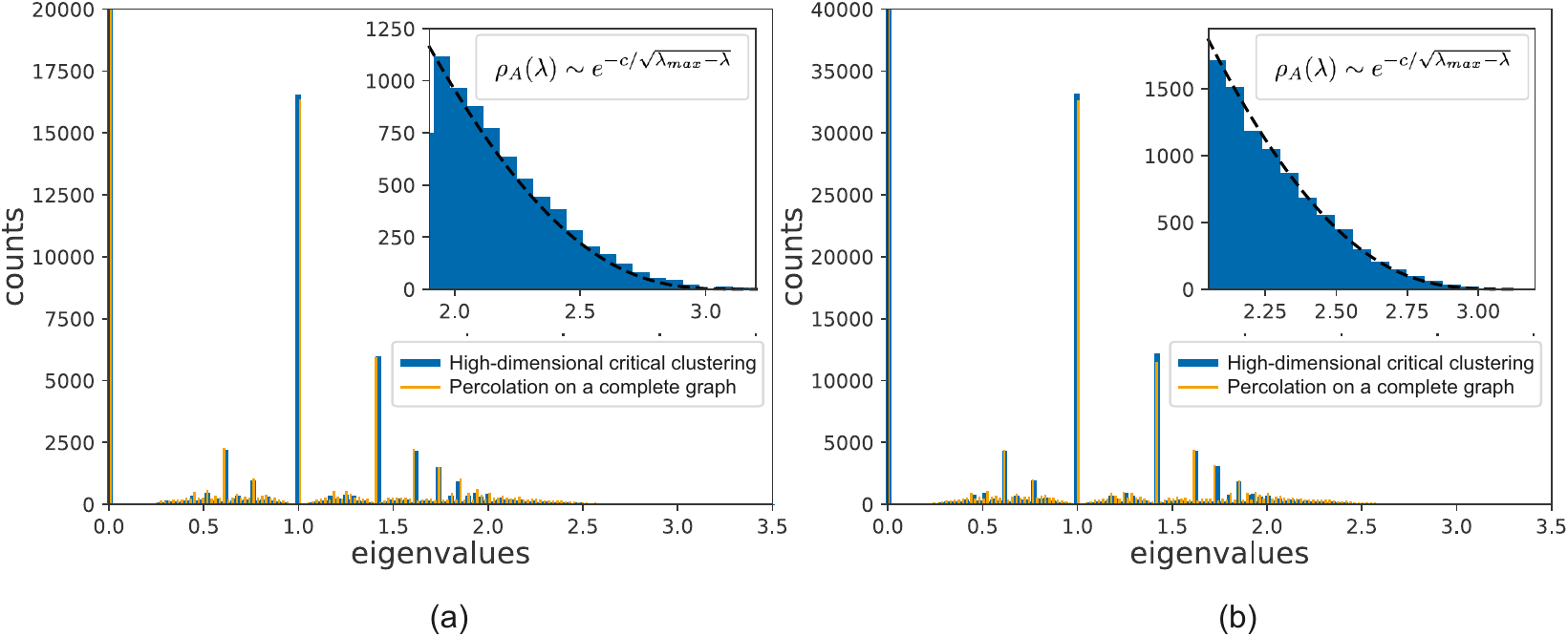}
\caption{Spectral densities of adjacency matrices for proximity RGG in a $D$-dimensional cube ($\rho_A(\lambda)$, blue) and ER graph constructed with the probability $p$ ($\rho_B(\lambda)$, orange) for two different sets of parameters $\{D,p\}=(200, 1/200)$ (a) and $\{D,p\}=(400, 1/400)$ (b).}
\label{fig:07}
\end{figure}

Our spectral analysis of datasets demonstrated very peculiar hierarchically organized pattern. The rare-event statistics naturally emerges in high dimensional spaces, where it manifests itself in the special hierarchical organization of distances between points, known as ``ultrametricity'' \cite{ultram}. Sparse statistics and ultrametricity together are rooted in high dimensionality and randomness. That has been unambiguously shown in \cite{zubarev}, where it was proved that in a $D$-dimensional Euclidean space the distances between points in sparse samplings tend to the ultrametric distances as $D\to\infty$.

Inserts in \fig{fig:06}a,b show the behavior of the tail of the spectral density of the adjacency matrix $\rho_A(\lambda)$ of RGG at $D_{cr}$. Numeric simulations are performed for $D_{cr}=220$. We found that enveloping curve of $\rho_A(\lambda)$ in the vicinity of the spectral boundary (i.e. near $\lambda_{max}$) matches the singular asymptotics known as the ``Lifshitz tail'' (LT):
\be
\rho_A(\lambda) = c_0 e^{-c/\sqrt{|\lambda_{max}-\lambda|}}
\label{eq:04}
\ee
We have chosen the parameters $c$ and $\lambda_{max}$ as optimization parameters to find the best fit of the enveloping shape of the spectral tail. We got $c=6.1$ and $\lambda_{max}=3.3$ (the insert in \fig{fig:06}a) and $c=5.9$ and $\lambda_{max}=3.26$ (the insert in \fig{fig:06}b). The obtained values of $\lambda_{max}$ are consistent with estimation of $\lambda_{max}$ from the maximal vertex degree $z_{max}=4$ discussed above. Let us point out that the emergence of LT at the percolation threshold in the spectral density tail of the Laplacian operator of ER graphs is known \cite{khorun, kirsch, avetisov} and its relation to the Gumbell statistics is the subject of recent studies \cite{bogdan}. 

\subsubsection{Spectral density of RGG above $D_{cr}$} 
\label{sect:3b-2}

We have also considered the spectral density $\rho_{A}(\lambda)$ of RGG above $D_{cr}$. For $D>D_{cr}$ the spectral density, $\rho_{A}(\lambda)$, consists of the main zone in a form of a Wigner semicircle, $\rho_W(\lambda)$, typical for the Gaussian matrix ensembles, with the boundary at $\lambda_{max}$, and one separated far-removed largest eigenvalue, $\Lambda_{max}$. The Wigner semicircle
\be
\rho_W(\lambda) = \frac{2}{\pi \lambda_{max}^2} \sqrt{\lambda_{max}^2-\lambda^2}
\label{eq:rho}
\ee
bounds the main zone of the spectrum by the values $\pm \lambda_{max}$, where 
\be
\lambda_{max} = f(D)\sqrt{N}
\label{eq:boundary}
\ee
and $f(D)$ is some function of $D$. The typical behavior of the spectral density $\rho_A(\lambda)$ in a dense regime, is shown in \fig{fig:07}a at $D=240$ for ensembles of adjacency matrices of RGG of size $N\times N$ (where $N=200$). 

\begin{figure}[ht]
\includegraphics[width=0.8\textwidth]{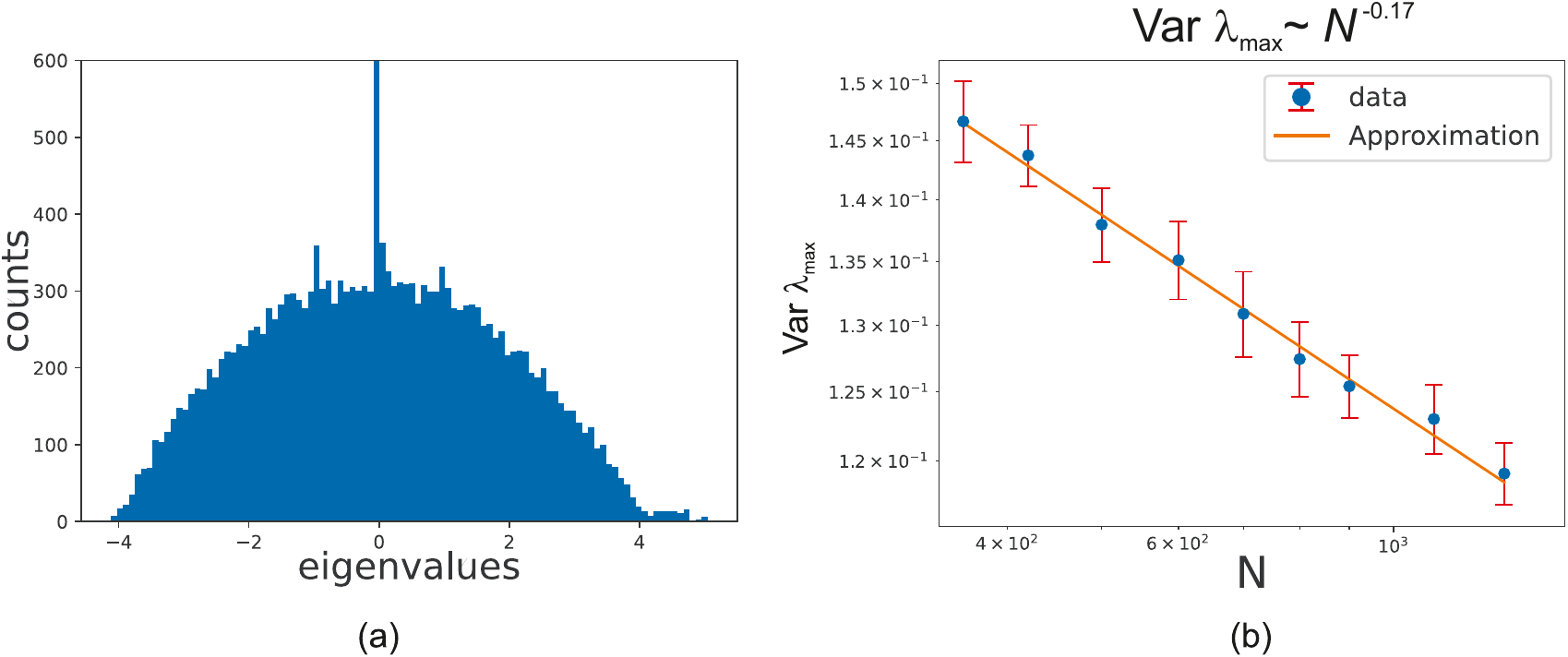}
\caption{(a) The spectral density $\rho_A(\lambda)$ of RGG in the dense regime for $D=240$; (b) The dependence of the variation of $\lambda_{max}$ on $N$.}
\label{fig:08}
\end{figure}

The behavior of the far-removed maximal eigenvalue, $\Lambda_{max}$, follows from the Perron-Frobenius theorem, which states that $\Lambda_{max}$ of a positive matrix $A=\{a_{ij}\}$ satisfies the bilateral inequality
\be
\min_i \sum_{j=1}^N a_{ij} \le \Lambda_{max} \le \max_i \sum_{j=1}^N a_{ij}
\label{eq:perron}
\ee
When $D\gg D_{cr}$ the graph becomes almost regular and $\Lambda_{max}$ is sandwiched between two variables, for both of them we approximately have $\bar{d} = \mathrm{const}\, D$ where $\bar{d}$ is the mean vertex degree of RGG. 

The estimate of the finite-size correction to the spectral boundary of the main zone, $\lambda_{max}(N)= f(D)\sqrt{N}$, in the dense regime can be done following the arguments of \cite{bowick}. Suppose that the function $f(D)$ depends on $D$ only and is $N$-independent. Set the typical distance, $\delta$, between adjacent eigenvalues in the vicinity of main zone boundary, $\lambda_{max}$. By definition the integral of $\rho(\lambda)$ over the interval $\left[\lambda_{max}-\delta, \lambda_{max}\right]$ is the fraction of eigenvalues falling within this range, i.e.
\be
\label{eq:lambda}
\int_{\lambda_{max}-\delta}^{\lambda_{max}} \rho(\lambda)\, d\lambda \simeq \frac{1}{N}
\ee
Plugging \eq{eq:rho} into \eq{eq:lambda} and taking into account that $\lambda_{max} = f(D)\sqrt{N}$, one arrives at the equation
\be
\frac{2}{\pi f^2(D)N} \int_{f(D)\sqrt{N}-\delta}^{f(D)\sqrt{N}} \sqrt{f^2(D)N-\lambda^2}\, d\lambda \simeq \frac{1}{N}.
\label{eq:lambda2}
\ee
which provides an estimate of subleading scaling correction, $\delta$ in the vicinity of $\lambda_{\rm max}$, valid at $N\gg 1$:
\be
\frac{4 \sqrt{2} \delta^{3/2}}{3 \pi f^{3/2}(D) N^{3/4}} \simeq \frac{1}{N}; \qquad \delta \simeq \frac{(3\pi)^{2/3}}{2^{5/3}} f(p) N^{-1/6}.
\label{lambda3}
\ee
Thus, the eigenvalue $\lambda_{max} \approx f(D)\sqrt{N}$ which bounds the continuous zone of the spectral density at large finite $N$ is defined with the uncertainty $\delta \sim f(D) N^{-1/6}$, i.e.
\be 
\lambda_{max} \simeq f(D)\sqrt{N} \pm  f(D) N^{-1/6}
\label{eq:lambda4}
\ee
To see how the fluctuations of the main zone boundary, $\lambda_{max}$, shrink with growing $N$, we have numerically computed the variance $\mathrm{Var}(\lambda_{max}) = \sqrt{\la\left(\lambda_{max}-\la \lambda_{max}\ra \right)^2 \ra}$ as a function of $N$. According to theoretical estimates following from \eq{eq:lambda}, one should have $\mathrm{Var}(\lambda_{max})\sim N^{-\xi}$ with $\xi=1/6\approx 0.167$. The result of our simulations is shown in \fig{fig:07}b from which one can extract $\xi \approx 0.17$ pretty close to the theoretical value $1/6$.

\section{Conclusion}
\label{sect:4}

Here, we have presented a model aimed at elucidating the phenomenon of sudden emergence of ``black days'', wherein various unrelated routine activities seemingly occur simultaneously. We depicted activities as $D$ independent periodic (characterized by a random period and a phase), or quasi-random sequences. By employing the Uniform Manifold Approximation and Projection (UMAP) technique, we observed a clustering of these activities in a two-dimensional manifold ${\cal M}$ when $D$ reaches a certain critical value, $D_{cr}$. This clustering signifies the manifestation of a ``black day''. In Appendix \ref{app1} we reconsider the same problem using the ``Laplacian Eigenmaps'' (LE) instead of UMAP and designate a class of problems where LE technique may offer advantages over UMAP.

The density of points on ${\cal M}$ serves as a quantitative measure of clustering in a proximity graph ${\cal G}$ within a $D$-dimensional space, representing all $D$ sequences of activities. Our simulations reveal that this density exhibits a saturation as the number of sequences is increasing below $D_{cr}$. Translating this observation to everyday life, we infer that within a certain range below $D_{cr}$, adding more new independent activities does not pose problems in activity management. Only upon reaching $D_{cr}$ one abruptly encounters a ``black day''.

The abstract concept of a "back day" can be understood in the context of traffic jams. By analyzing traffic flow using UMAP, we offer insights into the overall traffic patterns and the level of congestion across the road network under study. The RGG cutoff parameter $\kappa$ is easily interpretable in the context of traffic flow, as it characterizes the effective interaction between vehicles and can reflect typical road accessibility.

Examining the structure of a proximity $N$-edge graph ${\cal G}$ built on vertices of a $D$--dimensional cube ($N<D$) -- see Section \ref{sect:3} for details of construction -- we found that for fixed $N$ there is such $D_{cr}$ that the graph ${\cal G}$ represented by a collection of disconnected subgraphs, shrinks to a single-component cluster. Estimating the width of corresponding transition using finite-size scaling arguments, we conclude that it is of nearly third order. Above $D_{cr}$ the spectral statistics of ${\cal G}$ displays high similarity with the statistics of dense Erd\H{o}s-R\'{e}nyi graphs. The distinction between clustering in random regular graphs and in sparse matrices near the percolation threshold is discussed in Appendix \ref{app2}.

Our simulations (see \fig{fig:07}b) demonstrate that the variance of the fluctuations of the boundary of the main spectral zone for ensemble of finite random graphs of $N$ vertices in the $D$-dimensional space shrinks as $N^{-\xi}$ with $\xi\approx 0.17$, which is close to the expected theoretical value $\xi=1/6$. Our simulations do not permit to claim the existence of a Tracy-Widom distribution of $\lambda_{max}$ for RGGs, however provide hint for further research in this direction. 

To end up let us highlight the probabilistic paradox commonly referred to as the ``hot hand'', which presents itself as a seemingly unlikely sequence of successes, such as a prolonged series of goals in a basketball game. This phenomenon bears resemblance to the concepts discussed throughout our paper. Recently, Sidney Redner reevaluated the ``hot hand'' paradox in \cite{redner} through the lens of the backward Kolmogorov equation. While our perspective on the sudden accumulation of independent events differs fundamentally from the approach outlined in \cite{redner}, the phenomenon itself exhibits more similarities than differences.

\begin{acknowledgments}
We are grateful to P. Krapivsky, V. Avetisov and B. Slavov for insightful remarks and useful suggestions, and to L. Mirny and S. Maximov for valuable comments and relevant references. 
\end{acknowledgments}

\begin{appendix}

\section{Application of the Eigenmap Technique to the Data Clustering Analysis in Time Series Ensembles}
\label{app1}

The Eigenmap technique, often referred to as Laplacian Eigenmaps, is a nonlinear dimensionality reduction method that utilizes the eigenvalues and eigenvectors of the graph Laplacian matrix to embed high-dimensional data into a lower-dimensional (typically 2D) space \cite{belkin2001,belkin2003}. This technique aims to preserve the local neighborhood information of the data by constructing a weighted graph where nodes represent data points and edges signify their pairwise similarities. The resulting low-dimensional representation captures the intrinsic geometry of the data manifold, making it particularly useful for tasks such as clustering, visualization, and classification in complex datasets. 

\begin{figure}[ht]
\includegraphics[width=0.75\textwidth]{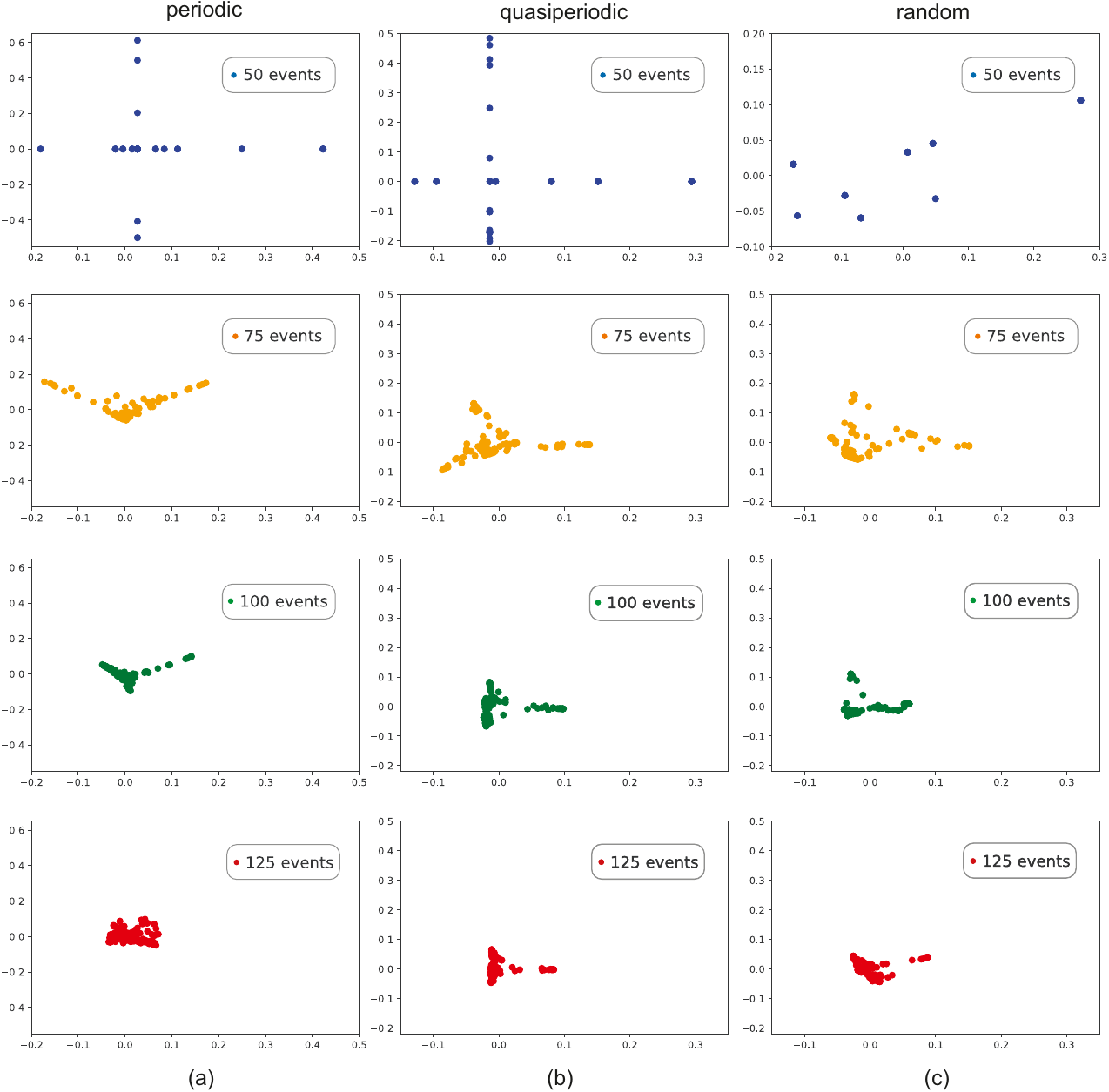}
\caption{Two-dimensional snapshots of activities discussed at length of Section \ref{sect:1} processed via the Eigenmap thechnique. The condensation of independent sequences of spikes for $D=\{50, 75, 100, 125\}$ is clearly seen with increasing $D$ for: periodic (a), quasiperiodic (b), and random (c) patterns. In the majority of realizations periodic and quasiperiodic sequences provide hairpin-like structures of clusters below the condensation threshold.}
\label{fig:09}
\end{figure}

UMAP and the Laplacian Eigenmaps are both nonlinear techniques that build a graph which represents the relationships between data points, emphasizing the preservation of local neighborhoods. However, UMAP extends this by optimizing a cross-entropy objective function to balance the local and global structure, aiming for a more faithful representation of the data's manifold structure. In contrast, Laplacian Eigenmaps rely on the spectral properties of the graph Laplacian matrix, using its eigenvalues and eigenvectors to perform the embedding. While Laplacian Eigenmaps are primarily based on spectral graph theory, UMAP incorporates elements from algebraic topology and Riemannian geometry, providing a more flexible and often more effective approach for visualizing and clustering complex datasets \cite{mcinnes2018umap}. Nevertheless, in some applications, the Eigenmap technique has advantages over UMAP. When applying Laplacian Eigenmaps to the analysis of time series, as discussed in detail in Section \ref{sect:1}, we obtain the results depicted in \fig{fig:09}. 

As seen, at low dimensionality, the Eigenmap pattern can distinguish periodic and quasiperiodic sequences from random ones: in the majority of realizations periodic and quasiperiodic sequences provide hairpin-like structures of clusters below the condensation threshold. However, above the critical dimension corresponding to data condensation, the difference between periodic and random sequences becomes invisible. It should be emphasized that observed difference between periodic and random sequences is statistical, and sometimes random sequences also produce regular patterns at low dimensionalities. Therefore, this issue requires more systematic investigation.

\section{Peculiarities of clustering in RGG and in sparse random matrices at percolation threshold}
\label{app2}

\begin{figure}[ht]
\includegraphics[width=0.75\textwidth]{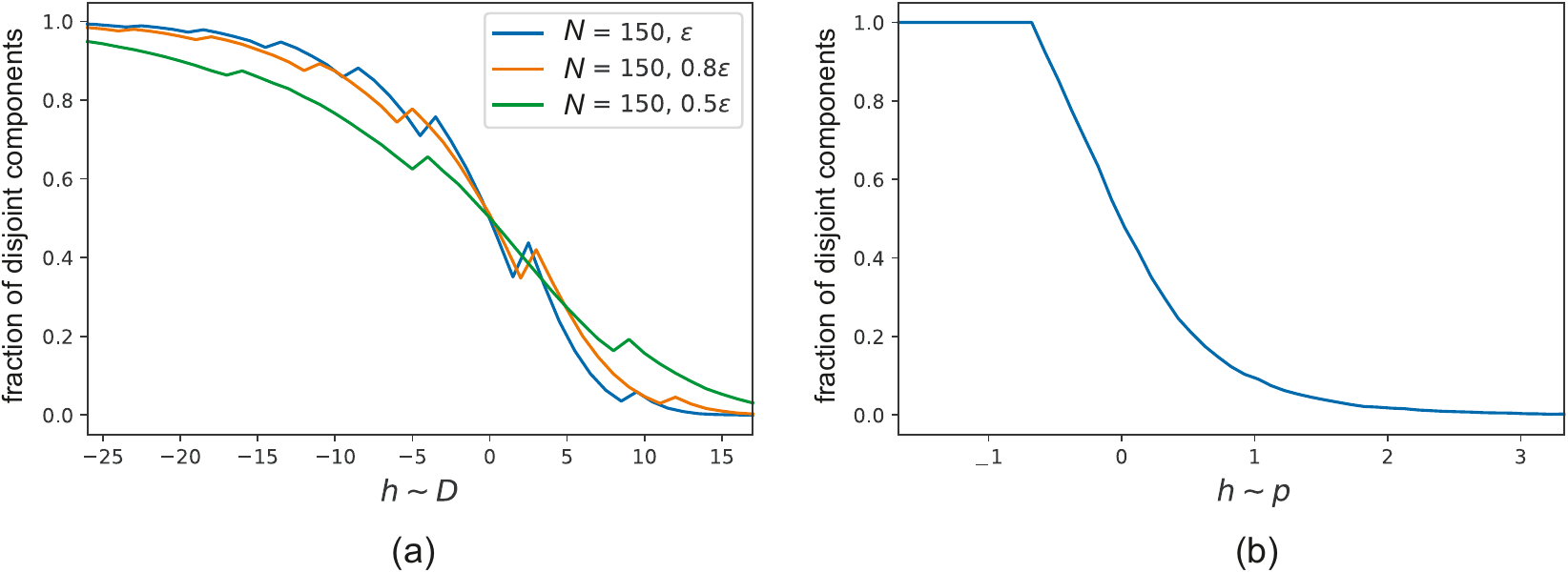}
\caption{(a) The dependence of the number of disjoint components in the multidimensional case for graphs with $N=150$ vertices and different thresholds $\varepsilon$, where $h=D - D_{cr}$; (b) the dependence of the number of disjoint components in the Erdős–Rényi graph with $N=150$ vertices where $h=n - n_{cr}$ where $n = 1 + pN$}
\label{fig:10}
\end{figure}

During the numerical simulations of phase transitions in multidimensional spaces, an interesting effect was observed: periodically, the number of connectivity components growth with increasing the space dimensionality, deviating from the general trend and creating a ``saw-teeth'' pattern, as shown in \fig{fig:10}a. In Section \ref{sect:3a}, we focused on the main trend in the dependence of the number of disjoint components as a function of $h$ and applied a smoothing filter in \fig{fig:04}a to smear up the saw-teeth-like behavior. However, we found the observed effect noteworthy and explore it in more detail here.

Computing the number of disjoint components as a function of the connection probability, $p$, in the standard Erd\H{o}s–R\'{e}nyi graph model, we have not observed the same ``saw-teeth effect'' for same graph sizes -- see \fig{fig:10}b. As $\eps$ decreases (which means that we effectively increase the dimensionality at which the phase transition occurs), the saw-teeth pattern gradually diminishes, and its frequency decreases, as shown in \fig{fig:10}a. Recall that the parameter $\eps$ defines the maximum distance between vectors in dimension $D$, where the corresponding graph vertices are considered to be adjacent. Thus, we hypothesize that the emergence of saw-teeth-like behavior in RGGs is the incommensurately effect linked to the discrete nature of the space dimensionality, $D$, which acts as the control parameter in the model.

\end{appendix}

\bibliography{main.bib}

\end{document}